\documentclass{iopart}
\begin{document}

\title[Model-independent definition and determination $\ldots$]{Model-independent definition and determination of electromagnetic formfactors and multipole moments}

\author{Frieder Kleefeld\dag  
}

\address{\dag\ Centro de F\'{\i}sica das Interac\c{c}\~{o}es Fundamentais (CFIF), Instituto Superior T\'{e}cnico (IST), Edif\'{\i}cio Ci\^{e}ncia, Av.\ Rovisco Pais, P-1049-001 Lisboa, Portugal}

\ead{kleefeld@cfif.ist.utl.pt}

\begin{abstract}
A theoretical method for the systematic definition and determination 
of Cartesian and spherical electromagnetic (onshell) formfactors and 
multipole moments for particles or composite systems from electromagnetic 
Breit-frame current distributions is presented. 
The method presented is free of sign ambiguities and is not based
on the underlying analytical substructure of the electromagnetic current distributions.
By construction the method contains all higher order momentum derivative
terms within the definition of electromagnetic formfactors and multipole 
moments, which are not taken into account in a lot of existing
theoretical calculations. 
\end{abstract}


\submitto{\jpg}

\maketitle

\section{Introduction} \label{sec1}

The experimental and theoretical investigation of electromagnetic properties of particles or composite systems like atomic nuclei has a long tradition (see e.g.\ \cite{gla1} and references therein), which is well known to be intimately related to the polarization properties of such systems (see e.g.\ \cite{ohl1,sta1} and references therein). The objective of the presented work is not to prove or disprove the impressive amount of previous work on this issue, yet to draw the attention of the physical community to two subtile points being hardly refered to in the vast list of present publications, yet yielding substantial quantitative uncertainties in the interpretation of theoretical results in comparison with experimental data. Throughout the work presented an unique prescription will be given how to circumvent these uncertainties, in order to come to an unambiguous notation and result.

The first point is the uncertainty in fixing or extracting the absolut sign of and the relative phase between different electromagnetic multipole moments of systems under consideration. The source of uncertainty  and mistakes is here on one hand the various conventions in notation appearing in literature, the different (spacelike or timelike Euclidean or Minkowski) metrics used\footnote{The metric tensor used in this publication is $g^{\mu\nu}=$ Diag$(+1,-1,-1,-1)$ with $\mu, \nu$ $\in$ $\{0,1,2,3\}$.}, the different defining relations for the four-momentum transfer $q$ and the absolut sign of the elementary charge unit $e$, which may lead to different results. On the other hand there is still a sizable uncertainty in relating Cartesian to spherical multipole moments depending on their definition by different authors. In polarization, i.e.\ spin physics this problem is partially cured by the so called Basel \cite{bas1} and Madison \cite{mad1} Convention, while in the description of electromagnetic current distributions a wide community of authors is still very ambiguous in their notations.

The second point is related to the fact that electromagnetic current distributions, formfactors and corresponding multipole moments are --- due to experimental purposes --- defined and experimentally detected in configuration space, while --- due to technical reasons --- in most cases calculated by theoreticians in momentum space. In configuration space the calculation of electromagnetic multipole moments yields integrals over the product of the configuration space current distribution and special combinations of space coordinates, while in momentum space --- after performing the respective Fourier transform and a partial integration --- the corresponding momentum integrals show up to contain momentum derivatives acting on the Fourier transform of the current distribution at places where in configuration space the space coordinates explicitely appeared (see e.g.\ \cite{hon1}). The problem is, that nowadays many theoreticians give or refer to defining equations of electromagnetic formfactors in momentum space, which are incomplete with respect to their exact definition in configuration space, as they ingore a gross part of the derivative terms mentioned above. Yet an exact definition of electromagnetic formfactors in momentum space is crucial to compare the various theoretical calculations present among each other and with experiment, especially at high momentum transfers, where results are most sensitive to the derivative terms mentioned.

In order to clarify the two points I want to consider already at this place the matrix element $<p^{\;\prime}\,;S^{\,\prime},S_z^{\,\prime}\;|J^\mu (x)|\,p\;;S,S_z>$ of the current density operator $J^\mu (x)$ between incoming and outgoing {\em onshell} state vectors of four-momentum $p$, $p^{\;\prime}$ and spin-quantum numbers $S,S_z$ and $S^{\,\prime},S_z^{\,\prime}$ for a vector particle ($S=1$) of mass $M$, which under restriction of translational invariance (upto an overall velocity dependent term $p+p^\prime$) is a function of the four-momentum transfer $q:= p^\prime - p$ and under additional restrictions of Lorentz-convariance and time-reversal invariance parametrized according to e.g.\ \cite{hon1,gou1,bro2,zui1,kim1} by three formfactors $F_1 (q^2)$, $F_2 (q^2)$ and $G_1 (q^2)$ (see also Section \ref{sec3} and \ref{appd1}): 
\begin{eqnarray} 
\bs\fl <p^{\;\prime}\,;1,S_z^{\,\prime}\;|J^\mu (0)|\,p\;;1,S_z> \quad = \nonumber \\
\bs \fl - \, e \, \left(\varepsilon^{\,\rho \, S_z^{\prime}} (\vec{p}^{\,\prime}\,)\right)^\ast \Bigg[ (p^\mu +p^{\,\prime \,\mu}) \, \left[ \, g_{\rho\sigma} \, F_1 (q^2) - \, \frac{q_\rho \,q_\sigma}{2\,M^2} \; F_2 (q^2) \,\right] \pm \delta^{\,\mu\nu}_{\rho\sigma}\, q_\nu \; G_1 (q^2) 
 \Bigg] \, \varepsilon^{\,\sigma \, S_z} (\vec{p}\,) \nonumber \\
\bs\fl \qquad\qquad ( \; \delta^{\,\mu\nu}_{\rho\sigma} \; := \;  
 g^{\,\mu}_{\,\rho} \, g^{\,\nu}_{\,\sigma} - g^{\,\mu}_{\,\sigma} \, g^{\,\nu}_{\,\rho} \, ,
  \quad \mbox{ $e$ := positive elementary charge unit}). \label{rel0}
\end{eqnarray} 
The polarization vectors $\varepsilon^{\,\mu \, S_z} (\vec{p}\,)$ with $S_z=0,+1,-1$  fulfil the following properties:
\begin{eqnarray} \bs \fl {\left( {\varepsilon}^{\,\mu \,S_z} (\vec{p}\, ) \right)}^\ast \, 
 {\varepsilon}_{\,\mu}^{\;\;\,S_z^\prime} (\vec{p}\, ) = - \,
{\delta}_{\,S_z S_z^\prime} \, , \quad
  \sum_{S_z} \, {\varepsilon}^{\,\mu \,S_z} (\vec{p}\, ) \, {({\varepsilon}^{\,\nu \,S_z} (\vec{p}\, ))}^{\ast}
  =  - \,
 g^{\,\mu \nu} + \frac{p^{\,\mu} \,p^{\,\nu}}{M^{\,2}} \;\; . \nonumber 
\end{eqnarray}
They are transverse ($p_{\,\mu} \; {\varepsilon}^{\,\mu \,S_z} (\vec{p}\, )|_{p^0 = \omega \,(|\vec{p}\,|)} = 0$ with $\omega \,(|\vec{p}\,|) :=\sqrt{\vec{p}^{\,2} + M^2}\,$), while the momentum states are normalized for arbitrary spin according to:
\begin{eqnarray} \bs \fl <p^{\;\prime}\,;S^{\,\prime},S_z^{\,\prime}\;|\,p\;;S,S_z> 
\,\; = \; 
(2\pi)^3 \; 2 \, \omega \,(|\vec{p}\,|) \; \delta^{\,3}(\vec{p}^{\,\prime}-\vec{p}\,\,) \;
\delta_{SS^{\prime}} \; \delta_{S_z S_z^{\prime}} \;\; . \label{nrm1} 
\end{eqnarray}
It is easy to see, that the introduction of the formfactors $F_1 (q^2)$, $F_2 (q^2)$ and $G_1 (q^2)$ is based on the idea, that the matrix element of the current distribution operator can be decomposed in polarization vectors and further analytical structures. Afterwards --- 
what is done in a lot of publications by now (see e.g.\ \cite{gou1,bro2,zui1}) --- many authors introduce for a vector  particle the electric charge formfactor $F_C(q^2)$, the magnetic dipole formfactor $F_M(q^2)$ and the electric quadrupole formfactor $F_Q(q^2)$ by performing the following identifications in momentum space ($\eta := - \, q^2 / (2\,M)^2$):
\begin{eqnarray}
\eqalign{F_C\,(q^2) & = \,F_1(q^2) + \frac{2}{3} \;\eta \, \left[ F_1(q^2) + (1 + \eta ) \, F_2\,(q^2) \,\pm \,G_1(q^2) \right] \\
\bs F_M(q^2) & \stackrel{?}{=} \, \pm \; G_1(q^2) \\
\bs F_Q\,(q^2) & \stackrel{?}{=} \,F_1(q^2) + (1 + \eta ) \, F_2\,(q^2) \,\pm\, G_1(q^2)\;\; . } \label{idad1} 
\end{eqnarray}
Another definition of $F_C(q^2)$,  $F_M(q^2)$ and $F_Q(q^2)$  based on reduced matrix elements of a momentum space current distribution is given in \cite{are1,are2}. A complementary review on the corresponding definition of electromagnetic formfactors for systems with arbitrary spin can be found in e.g.\ \cite{akh1}. It should be mentioned that in  \cite{akh1} the definition of magnetic formfactors is based on the static magnetization vector $\vec{\mu} (\vec{r}\,)$ and not on the static current distribution vector $\vec{j} (\vec{r}\,)$.  The identification $\vec{j} (\vec{r}\,) =\vec{\nabla} \times \vec{\mu} \, (\vec{r}\,)\,$ is consistent with the static continuity equation $\vec{\nabla} \cdot \vec{j} \, (\vec{r}\,) = 0$. 

The electromagnetic formfactors $F_C(q^2)$,  $F_M(q^2)$ and $F_Q(q^2)$ are defined such, that $e\,F_C(0)$,  $\frac{e}{M} \, F_M(0)$ and $\frac{e}{M^2} \, F_Q(0)$ are the vector particle's electric charge, the magnetic dipole moment and the electric quadrupole moment respectively. That this is not true for at least the identifications of the magnetic and the quadrupole formfactors given in (\ref{idad1}) will be shown in the following text. As has been discussed above the reason is that the identifying expressions (\ref{idad1}) for $F_M\,(q^2)$ and $F_Q\,(q^2)$ don't contain contributions of small, yet relevant terms with momentum derivatives of the formfactors $F_1\,(q^2)$, $F_2\,(q^2)$ and $G_1(q^2)$. In several calculations such contributions give finite results even for $q^2=0$, which are part of the so called {\em intrinsic contributions} to the electromagnetic multipole moments of the vector particle \cite{hon2}. As will be shown in the case of the charge formfactor no derivatives appear, i.e. the identification for $F_C\,(q^2)$ given in (\ref{idad1}) is exact and the charge does not get any further correction from an intrinsic contribution. 

The sign ambiguity in (\ref{rel0}) denoted by $\pm \; G_1(q^2)$ (affected by the choice of the sign of elementary charge unit $e$) causes to several authors problems, who don't know how to fix the sign properly. In general authors make the identification $F_M(q^2) \stackrel{!}{=} +\, G_1(q^2)$, yet choose the signs in front of  $G_1(q^2)$ in the expressions for $F_C\,(q^2)$ and $F_Q\,(q^2)$ according to formulae taken from different articles quoted --- often negative, even for positive unit charge. In the comparison to experiment --- say in elastic electron-deuteron scattering --- this sign ambiguity seem to play a marginal role, especially for very small momentum transfers. E.g.\ the well known Rosenbluth formfactors $A(q^2)$ and $B(q^2)$, which usually are identified by (see e.g.\ \cite{zui1,are1,gar1} and references therein)
\begin{eqnarray} \fl A(q^2) \stackrel{?}{=} F_C^2 (q^2) + \frac{8}{9} \,{\eta}^2\,F_Q^2 (q^2) + \frac{2}{3} \,\eta \, F_M^2 (q^2) \, , \quad 
 B(q^2) \stackrel{?}{=} \frac{4}{3} \, \eta \,(\eta + 1)\:F_M^2 (q^2)\; , \label{idad2}
\end{eqnarray}
depend only on the square of $F_M (q^2)$, which is insensitive to the sign of $G_1(q^2)$. Yet at least the formfactor $A(q^2)$ depends also on $\eta^2\, F^2_Q(q^2)$, which is sensitive to the sign of $G_1(q^2)$ for $q^2\not= 0$. In the case of the deuteron for small $q^2$ the quadrupole formfactor is dominated by $F_2(q^2)$, as there holds approximately the following proportionality $|F_1(0)| : |F_2(0)| :|G_1(0)|  \simeq 1 : 25 : 1.7$ (see e.g.\ \cite{gar1}), yet for larger negative values of $q^2$ their relative weights change drastically, so a sign change or wrong sign in front of $G_1(q^2)$  or intrinsic contributions to electromagnetic formfactors can be quantitatively very relevant. The same is of course true for the tensor polarization $T_{20} (q^2)$, which is usually introduced by the following identification \cite{are1,are2,gar1}:
\begin{eqnarray} \lo T_{\,20} (q^2)  \stackrel{?}{=} - \, \sqrt{2} \,\,\; \frac{x (x+2) + y/2}{1+2\,(x^2+y)} \label{idad3} 
\end{eqnarray}
with
\begin{eqnarray} \lo x:= \frac{2 \,\eta \, F_Q (q^2)}{3 \, F_C (q^2)}\; , \quad y:= \frac{2 \,\eta}{3} \,\left(\,\frac{1}{2} + (1 + \eta ) \, \tan^2\, \frac{{\theta}_e}{2}  \right) \,\, {\left( \frac{F_M (q^2)}{F_C (q^2)} \right) }^2 \; . \nonumber 
\end{eqnarray}
The reason, why a questionmark is put on all identifications of $A(q^2)$, $B(q^2)$, $T_{20} (q^2)$ with $F_C (q^2)$, $F_M (q^2)$, $F_Q (q^2)$ (equations (\ref{idad2}) and (\ref{idad3})) is, that they all are derived by comparison of expressions for the differential cross section of elastic electron-deuteron scattering containing $A(q^2)$, $B(q^2)$, $T_{20} (q^2)$ with theoretical results derived via the matrix element (\ref{rel0}) of the current distribution operator given in terms of $F_1 (q^2)$, $F_2 (q^2)$, $G_1 (q^2)$ and later reexpressed by $F_C (q^2)$, $F_M (q^2)$, $F_Q (q^2)$ given by the inexact identifying equations (\ref{idad1}) discussed above. Future calculations will have to take this issue with much greater care.

The goal of this publication is to give an unambiguous exact definition in configuration space and derivation in momentum space of the (onshell) electromagnetic formfactors of particles or composite systems with arbitrary spin in terms from the Breit-frame matrix elements of the current-distribution operator, without knowledge, whether the matrix element of the current-distribution operator can be decomposed into polarization vectors or not. A first application of this formalism with respect to the deuteron within a Bethe-Salpeter framework is already available \cite{kle1}. It turns out, that intrinsic contributions to electromagnetic formfactors have sizable effects.

The paper is organised as follows: in Section \ref{sec2} the electromagnetic multipole moments of a classical current distribution in configuration space and momentum space are defined and reconsidered; in Section \ref{sec3} after extending the formalism within the Breit-frame to a current distribution in Quantum-Field Theory (QFT) consistent expressions for the electric and magnetic formfactors and multipole moments in QFT are derived; Section \ref{sec4} is closing with a summary and a short outlook. 

\section{Electromagnetic current distributions and their multipole moments} \label{sec2} 
\subsection{Spherical multipole moments defined in configuration space}
The electrostatic and magnetostatic potentials $\Phi_E (\vec{r}\,)$ and 
$\Phi_M (\vec{r}\,)$ and the corresponding spherical electric and magnetic
multipole moments $E_{\ell m}$ and $M_{\ell m}$ are according to \cite{bro1} connected to the
static electromagnetic current distribution $j^\mu (\vec{r}\,)$ and the electric
and magnetic fields $\vec{E} (\vec{r}\,)$ and $\vec{B} (\vec{r}\,)$ in the 
following way ($r := |\vec{r}\,| $):
\begin{eqnarray}
\fl \eqalign{\Phi_E (\vec{r}\,) \; = \;
- \,\int^{\;\displaystyle\vec{r}}_{\infty}d\vec{\xi} \cdot \vec{E} (\vec{\xi}\,) & \; = \;
\sum\limits_{\ell m} \frac{4\pi}{2\ell +1}
 \; \frac{E^{\,\ell m}}{r^{\ell + 1}} \; Y_{\ell m} (\Omega) 
\; = \; 
\int d^3r^{\;\prime} \;\; \frac{j^{\,0}(\vec{r}^{\,\prime})}{
\left| \vec{r} - \vec{r}^{\;\prime}\right|} \\
\Phi_M (\vec{r}\,) \; = \;
- \, \int^{\;\displaystyle\vec{r}}_{\infty}d\vec{\xi} \cdot \vec{B} (\vec{\xi}\,) & \; = \;
\sum\limits_{\ell m} \frac{4\pi}{2\ell +1}
 \; \frac{M^{\,\ell m}}{r^{\ell + 1}} \; Y_{\ell m} (\Omega) \\
 & \; = \; \int_{\infty}^{\;\displaystyle r} \frac{d\rho}{\rho}
\int d^3r^{\;\prime} \;\; \frac{\vec{\nabla}^{\prime} \cdot 
\left[ \vec{r}^{\;\prime} \times \vec{j} (\vec{r}^{\;\prime}) \right]}{
\left| \rho \; \frac{\displaystyle\vec{r}}{\displaystyle r} - \vec{r}^{\;\prime}\right|} \; .} \nonumber 
\end{eqnarray}
The sum $\sum_{\ell m}$ yields of course $\ell = 0,1,2,\ldots, \infty$ and $m = -\,\ell,-\ell +1, \ldots, \ell$.
It is now straight forward to establish the following relations between the spherical electric and magnetic multipole moments $E_{\ell m}$ and $M_{\ell m}$ and the current distribution $j^\mu (\vec{r}\,)$ in configuration space:
\begin{eqnarray} 
\eqalign{E^{\,\ell m} & = \; \int d^3r \; r^\ell \; Y^\ast_{\ell m} (\Omega) \; 
j^{\,0}(\vec{r}\,)  \\
M^{\,\ell m} & = \; \int d^3r \; r^\ell \; Y^\ast_{\ell m} (\Omega) \; 
\frac{\vec{\nabla} \cdot 
\left[ \vec{j} (\vec{r}\,) \times \vec{r} \right]}{\ell + 1} \; .} \label{rel1}   
\end{eqnarray}
Without loss of generality the solid angles can be chosen with respect to the $z$- and $x$-direction. To perform a unique relation between spherical and Cartesian electromagnetic formfactors and multipole moments its now crucial to consider the following Cartesian decomposition of the solid
harmonics $r^\ell\,Y_{\ell m} (\Omega)$ (see e.g.\ \cite[p.\ 133]{var1}):
\begin{eqnarray} \fl \eqalign{\bs r^{\,\ell}\,Y^\ast_{\ell m}(\Omega) & = \;
\sqrt{\frac{2\ell +1}{4\pi}\, (\ell+m)! (\ell-m)!} \;\; 
\sum\limits_{p\,q\,n} \frac{1}{p\,!\,q\,!\,n\,!} 
{\left( - \, \frac{x-i\,y}{2} \right)}^p
{\left( \frac{x+i\,y}{2} \right)}^q
z^n \\
\bs & =: \; \sqrt{\frac{2\ell +1}{4\pi}} \; \; \frac{b^{\,\ell m} (\vec{r}\, )}{\ell
 \,!} \; =: \; 
 \sqrt{\frac{2\ell +1}{4\pi}} \; \; \frac{(b_{\,\ell m} (\vec{r}\, ))^\ast}{\ell
 \,!}  \\
\bs &  
 \mbox{($p,q,n$ = all positive integers with $p+q+n=\ell$ and $p-q=m$)} \; .} 
\end{eqnarray}
The decomposition has been used to define the polynomials $b^{\,\ell m} (\vec{r}\, ) = (b_{\,\ell m} (\vec{r}\, ))^\ast$. For $\ell = 0,1,2$ they are given by:
\begin{eqnarray}
\fl \eqalign{b^{\,00} (\vec{r}\,) & \; = \;\; (b_{\,00} (\vec{r}\,))^\ast \;\;\; = \; 1  \\
b^{\,11} (\vec{r}\,) & \; = \; \; (b_{\,11} (\vec{r}\,))^\ast \;\;\; 
= \; - (x -i\,y)/\sqrt{2} \\
b^{\,10} (\vec{r}\,) & \; = \; \; (b_{\,10} (\vec{r}\,))^\ast \;\;\; = \; z  \\
b^{\,1-1} (\vec{r}\,) & \; = \; \; (b_{\,1-1} (\vec{r}\,))^\ast \; 
=  + (x +i\,y)/\sqrt{2} \\
b^{\,22} (\vec{r}\,) & \; = \; \; (b_{\,22} (\vec{r}\,))^\ast \;\;\;
= \;\sqrt{3/2} \; (x-i\,y)^2 \; \stackrel{!}{=} \; \sqrt{3/2} \; (x^2 - y^2 - 2\,
i\, x\, y ) \\
b^{\,21} (\vec{r}\,) & \; = \; \; (b_{\,21} (\vec{r}\,))^\ast \; \;\;
= \; - \, \sqrt{6} \; (x -i\,y) \, z  \\
b^{\,20} (\vec{r}\,) & \; = \; \; (b_{\,20} (\vec{r}\,))^\ast \; \;\;
= \; 2 \, z^2 - x^2 - y^2 \;\;\;\, \stackrel{!}{=} \; 3 \, z^2 - \vec{r}^{\;2} \quad  \\
b^{\,2-1} (\vec{r}\,) & \; = \; \; (b_{\,2-1} (\vec{r}\,))^\ast \; 
= \; + \, \sqrt{6} \; (x +i\,y) \, z  \\
b^{\,2-2} (\vec{r}\,) & \; = \; \; (b_{\,2-2} (\vec{r}\,))^\ast \; 
= \; \sqrt{3/2} \; (x+i\,y)^2 \; \stackrel{!}{=} \; \sqrt{3/2} \; (x^2
- y^2 + 2\,
i\, x\, y ) \; .} \nonumber 
\end{eqnarray}
The polynomials $b^{\,\ell m} (\vec{r}\, )$ fulfil the following useful property:
\begin{eqnarray} b^{\,\ell \,m} (\frac{\partial}{\partial\,\vec{r}}\, ) \;\, 
b_{\,\ell \,m^{\,\prime}} (\vec{r}\, ) \; = \; \frac{\ell\,!\,(2\,\ell
\,)!}{2^{\,\ell}} \;\; \delta_{m\,m^{\,\prime}\;\;.} \label{usf1} 
\end{eqnarray}
Using the polynomials $b^{\,\ell m} (\vec{r}\, )$ relation (\ref{rel1}) between the 
spherical electric and magnetic multipole moments $E_{\ell m}$ and $M_{\ell m}$ and the configuration space current distribution $j^\mu (\vec{r}\,)$ can be reformulated:
\begin{eqnarray} 
\eqalign{E^{\,\ell m} & = \; \frac{1}{\ell\, !} \; \sqrt{\frac{2\,\ell +1 }{4\pi}} \; \; \int d^3r \; b^{\,\ell m} (\vec{r}\, ) \; 
j^{\,0}(\vec{r}\,) \\
M^{\,\ell m} & = \; \frac{1}{\ell\, !} \; \sqrt{\frac{2\,\ell +1 }{4\pi}} \; \; \int d^3r \; b^{\,\ell m} (\vec{r}\, ) \; 
\frac{\vec{\nabla} \cdot 
\left[ \vec{j} (\vec{r}\,) \times \vec{r} \right]}{\ell + 1} \; .} \label{rel2}   
\end{eqnarray}
\subsection{Cartesian multipole moments defined in configuration space}
To obtain the Cartesian multipole moments one has to perform a
Taylor-expansion of $1/|\vec{r} - \vec{r}^{\;\prime}|$ in the variable 
$\vec{r}^{\;\prime}$, i.e.:
\begin{eqnarray} 
\fl \frac{1}{\left|\vec{r} - \vec{r}^{\;\prime}\right|} \; = \;
\frac{1}{r} +
\frac{1}{1!} \; r^{\,\prime \, i} \; \frac{c^{\,i}(\vec{r}\,)}{r^3} \; + \;
\frac{1}{2!} \; r^{\,\prime \, i} \; r^{\,\prime \, j} \;\frac{c^{\,ij}(\vec{r}\,)}{r^5} \; + \;
\frac{1}{3!} \; r^{\,\prime \, i} \; r^{\,\prime \, j} \; r^{\,\prime \, k} 
\;\frac{c^{\,ijk}(\vec{r}\,)}{r^7} \; + \; \ldots \; ,\nonumber
\end{eqnarray}
while the numerators of the expansion coefficients are polynomials given by:
\begin{eqnarray}
\eqalign{c^{\,i} (\vec{r}\,) & = \; r^{\, i}  \\
c^{\,ij} (\vec{r}\,) & = \; 3\, r^{\, i} r^{\, j} - \vec{r}^{\;2} \, \delta^{\,ij}  \\
c^{\,ijk} (\vec{r}\,) & = \; 15\, r^{\, i} r^{\, j} r^{\, k} 
- 3 \,\vec{r}^{\;2} \, ( 
r^{\, i} \, \delta^{\,jk} + r^{\, j} \, \delta^{\,ki} + r^{\, k} \, \delta^{\,ij}) \\
 & \cdots} \nonumber
\end{eqnarray}
Using these expansion coefficients the Cartesian electric and magnetic multipole
moments $Q_E$ and $Q_M$ of the current distribution $j^\mu (\vec{r}\,)$ can --- in correspondence to (\ref{rel2}) --- be
uniquely introduced by:
\begin{eqnarray} 
\eqalign{Q_E^{\;\displaystyle i_1\ldots i_\ell} & := \; \int d^3r \;\; c^{\;\displaystyle
i_1\ldots i_\ell} (\vec{r}\,) 
\; j^{\,0}(\vec{r}\,) \\
Q_M^{\;\displaystyle i_1\ldots i_\ell} & := \; \int d^3r \;\; c^{\;\displaystyle
i_1\ldots i_\ell} (\vec{r}\,) 
\; \frac{\vec{\nabla} \cdot 
\left[ \vec{j} (\vec{r}\,) \times \vec{r} \right]}{\ell + 1} \; .} \label{rel2a}  
\end{eqnarray}
It is easy to prove the following relations between the $m=0$ sherical and the corresponding Cartesian polynomials and multipole moments:
\begin{eqnarray} 
\eqalign{\ms c^{\;\displaystyle \overbrace{z\ldots z}^{\mbox{\footnotesize $\ell\,$ times}}} 
(\vec{r}\, ) & = \; 
 b^{\,\ell\,0} (\vec{r}\, ) \\
\ms Q_E^{\;\displaystyle z\ldots z} \quad \;\; & = \; 
\ell \,! \; \sqrt{\frac{4\pi}{2\ell +1}} \; E^{\,\ell\,0} \\
\ms Q_M^{\;\displaystyle z\ldots z} \quad \;\; & = \; 
\ell \,! \; \sqrt{\frac{4\pi}{2\ell +1}} \; M^{\,\ell\,0} \; .} \label{rel3} 
\end{eqnarray}
\subsection{Multipole moments defined in momentum space}
To obtain the corresponding definitions of the electromagnetic multipole moments in momentum space one has to perfor a Fourier transform of the static electromagnetic current distribution, i.e.:
\begin{eqnarray} j^{\,\mu} (\vec{r}\,) \; = \; \int \frac{d^3q}{(2\pi )^3} \;\; 
e^{\displaystyle -i\,\vec{q} \cdot \vec{r}} \; j^{\,\mu} (\vec{q}\,) \; . \label{rel4} 
\end{eqnarray}
In terms of the Fourier transform of a current distribution vanishing at infinity the Cartesian multipole moments (\ref{rel2a}) are determined by:
\begin{eqnarray} 
\fl \eqalign{Q_E^{\;\displaystyle i_1\ldots i_\ell} & = \; \int d^3q \;\; \delta^3(-\vec{q}\,) \;\;
c^{\;\displaystyle i_1\ldots i_\ell} (-i\,\frac{\partial}{\partial\vec{q}}\;) 
\; j^{\,0}(\vec{q}\,) \\
Q_M^{\;\displaystyle i_1\ldots i_\ell} & = \; \int d^3q \;\; \delta^3(-\vec{q}\,) \;\;
c^{\;\displaystyle i_1\ldots i_\ell} (-i\,\frac{\partial}{\partial\vec{q}}\;) 
\; \frac{ \displaystyle\frac{\partial}{\partial\vec{q}} \cdot 
\left[ \vec{j} (\vec{q}\,) \times \vec{q} \, \right]}{\ell + 1} \; .} \label{rel5}  
\end{eqnarray}
The corresponding spherical multipole moments (\ref{rel2}) are:
\begin{eqnarray} 
\fl \eqalign{E^{\, \ell m} & = \; \frac{1}{\ell\, !} \; \sqrt{\frac{2\,\ell +1 }{4\pi}} \; \int d^3q \;\; \delta^3(-\vec{q}\,) \;\;
b^{\, \ell m} (-i\,\frac{\partial}{\partial\vec{q}}\;) 
\; j^{\,0}(\vec{q}\,) \\
M^{\, \ell m} & = \; \frac{1}{\ell\, !} \; \sqrt{\frac{2\,\ell +1 }{4\pi}} \; \int d^3q \;\; \delta^3(-\vec{q}\,) \;\;
b^{\, \ell m} (-i\,\frac{\partial}{\partial\vec{q}}\;) 
\; \frac{ \displaystyle\frac{\partial}{\partial\vec{q}} \cdot 
\left[ \vec{j} (\vec{q}\,) \times \vec{q} \,\right]}{\ell + 1} \; .} \label{rel6}   
\end{eqnarray}
\section{Multipole moments and formfactors in Quantum-Field Theory (QFT)} \label{sec3}
\subsection{Construction of current distribution operators in QFT}
The photon part of the Lagrange density of a particle (electromagnetic current distribution $j^{\,\mu} (x)\,$) interacting with an electromagnetic vector field $A^{\,\mu} (x)$ (electromagnetic field strength tensor \,$ F^{\,\mu\nu} \; = \; \partial^{\,\mu} A^{\,\nu} \, - \, \partial^{\,\nu} A^{\,\mu}$\,) is given by:
\begin{eqnarray} 
{\cal L}_{em} (x) \; = \; - \,\, A^{\,\mu} (x) \; j_{\,\mu} (x) \, - \, \frac{1}{4} \, F^{\,\mu\nu} \, F_{\,\mu\nu}  \, + \, {\cal L}_{\, gauge} \, (x) \; .\nonumber
\end{eqnarray}
In the covariant gauge the gauge-fixing Lagrangian is ${\cal L}_{\, gauge} \, (x) \, = \, - \, \zeta \; (\,\partial_\nu A^{\,\nu} \,)^{\,2}/2\,$.
The classical inhomogeneous Maxwell equation obtained by the variation of the action with respect to the vector field $A^{\,\mu} (x)$ is:
\begin{eqnarray}  (\,g^{\,\mu\nu} \, \opensquare_x  \,-\, (1\,-\,\zeta )\; \partial_x^{\,\mu} \,\partial_x^{\,\nu}  \,) \; A_{\,\nu} (x) \;=\; j^{\,\mu} (x) \; . \nonumber
\end{eqnarray}
The photon propagator for the photon in covariant gauge is given by:
\begin{eqnarray} 
\fl <0|T[\,A^{\,\mu} (x) \, A^{\,\nu} (y) \,]|0> \; = \; 
\int \frac{d^4k}{(2\,\pi)^4} \; \;e^{-ik\cdot (x-y)} \; 
i \, \left( \frac{-\, g^{\,\mu\nu}}{k^2 + i\, \varepsilon}
 \; + \;
\frac{\zeta - 1}{\zeta} \; \frac{k^{\,\mu}k^{\,\nu}}{(k^2 + i\, \varepsilon )^{\,2}}
\right) \; .\nonumber 
\end{eqnarray}
Choosing the covariant gauge therefore the electromagnetic current operator consistently can be defined as:
\begin{eqnarray} J^{\,\mu} (x) \; := \; 
(\,g^{\,\mu\nu} \, \opensquare_x  \,-\, (1\,-\,\zeta )\; \partial_x^{\,\mu} \,\partial_x^{\,\nu}  \,) \;
T \left[  \,  A_{\,\nu} (x) \; e^{\displaystyle\,i:\hat{{\cal S}}_{int}:}  \,\right] \; . \nonumber 
\end{eqnarray}
The same defining procedure works for any other gauge. Although the approach seems to be quite gauge dependent the resulting current distribution operator does not carry any external gauge dependence, if one uses for internal contractions the photon propagator for the selected gauge. Surely, from a field theoretic point of view, it is well known the current operators are singular objects and need some regularization (like point splitting etc.). But this feature doesn't affect the main results discussed in the following sections.

By the way --- the ansatz above yields the same results as one would obtain due to the (transverse) current operators being discussed in \cite{gro5}.
\subsection{Matrix elements of the current distribution operator and the Breit-frame}
In order to connect classical observable quantities like an electromagnetic current distribution $j^{\,\mu} (x)$ which Quantum Mechanical or Quantum-Field Theoretical operators like e.g.\ $J^{\,\mu} (x)$ one has to perform an expectation value of the respective operator with respect to state vectors decribing status of the system.
 
The matrix element of the current-distribution operator $J^\mu (x)$ between incoming and outgoing
(onshell) state vectors of four-momentum $p$, $p^{\;\prime}$ and spin-quantum
numbers $S,S_z$ and $S^{\,\prime},S_z^{\,\prime}$ (describing respective incoming and outgoing particles or composite systems) with normalization (\ref{nrm1}) has been introduced by:
\begin{eqnarray} \fl < p^{\;\prime}\,;S^{\,\prime},S_z^{\,\prime}\;|J^\mu (x)|\,p\;;S,S_z> 
 \; = \; e^{\displaystyle i\, q \cdot x}
 < p^{\;\prime}\,;S^{\,\prime},S_z^{\,\prime}\;|J^\mu (0)|\,p\;;S,S_z> \; . 
\end{eqnarray}
$q$ is the four-momentum transfer defined by $q:=p^{\,\prime}-p$.
In order to relate the electro- and magneto-static results obtained in Section \ref{sec2} with corresponding matrix elements obtained in QFT, one has to consider the matrix elements of quantum operators in a frame of reference, in which they are static, i.e. time-independent.

In the case of the matrix element of the current-distribution operator the respective frame of reference is the Breit-frame, in which the energy transfer $q^{\,0}$ is zero, i.e. 
$q^{\,\mu} = (0,\vec{q}\,)$. In the Breit-frame the incoming and outgoing four-momenta are determined by
one universal three-momentum $\vec{k}$, i.e $p^{\,\prime \,\mu} = : (\omega (|\vec{k}\,|),\vec{k}\,)$ and $p^{\,\mu} = : (\omega (|\vec{k}\,|),-\,\vec{k}\,)$. It is straight forward to define in the Breit-frame ($B=$ ``Breit'') the matrix element of the current distribution operator by (see also \ref{appd1}):  
\begin{eqnarray}
\fl j_B^{\,\mu} (\vec{q}\,;S\,;S^{\,\prime}_z,S_z) \; := \;
\frac{1}{2\,\omega (|\vec{k}\,|)} \;
<\omega (|\vec{k}\,|),\vec{k} \,;S,S_z^{\,\prime}\;|J^{\,\mu} (0)|\,\omega (|\vec{k}\,|),-\,\vec{k}\;;S,S_z> \; .
\end{eqnarray}
The ``diagonal'' quantity $j_B^{\,\mu} (\vec{q};S) \; := \; j_B^{\,\mu}
(\vec{q}\,;S\,;S,S)$ plays the role of the classical observable Fourier transform of a current distribution of a particle or a composite system with spin $S$, i.e.\ after the Fourier transform (see equation (\ref{rel4}))
\begin{eqnarray} j_B^{\,\mu} (\vec{r}\,;S) \; := \; \int \frac{d^3q}{(2\pi )^3} \; \; 
e^{\displaystyle -\,i\,\vec{q} \cdot \vec{r}} \; j_B^{\,\mu} (\vec{q}\,;S) 
\end{eqnarray}
the quantity $j_B^{\,\mu} (\vec{r}\,;S)$ can be identified with the static current distribution $j^{\,\mu} (\vec{r}\,)$ discussed in Section \ref{sec2}.
\subsection{Momentum space definition of electromagnetic multipole moments in QFT} \label{sbsec1}
It is now straight forward to define according to (\ref{rel5}) the Cartesian electric and magnetic 
multipole moments $Q_E^{\;\displaystyle i_1\ldots i_\ell} (S)$
and $Q_M^{\;\displaystyle i_1\ldots i_\ell} (S)$
of a spin-$S$ particle or composite system using its Breit-frame current by:
\begin{eqnarray} \fl
Q_E^{\;\displaystyle i_1\ldots i_\ell} (S) \; := \;
Q_E^{\;\displaystyle i_1\ldots i_\ell} (S;S,S) \; , \quad
Q_M^{\;\displaystyle i_1\ldots i_\ell} (S) \; := \;
Q_M^{\;\displaystyle i_1\ldots i_\ell} (S;S,S) 
\end{eqnarray}
with 
\begin{eqnarray} \fl \eqalign{Q_E^{\;\displaystyle i_1\ldots i_\ell} (S;S^{\,\prime}_z,S_z) & := \; \int d^3q \;\; \delta^3(-\vec{q}\,) \;\;
c^{\;\displaystyle i_1\ldots i_\ell} (-i\,\frac{\partial}{\partial\vec{q}}\;) 
\; j_B^{\,0}(\vec{q}\,;S;S^{\,\prime}_z,S_z) \\
Q_M^{\;\displaystyle i_1\ldots i_\ell} (S;S^{\,\prime}_z,S_z) & := \; \int d^3q \;\; \delta^3(-\vec{q}\,) \;\;
c^{\;\displaystyle i_1\ldots i_\ell} (-i\,\frac{\partial}{\partial\vec{q}}\;) 
\; \frac{ \displaystyle\frac{\partial}{\partial\vec{q}} \cdot 
\left[ \vec{j}_B (\vec{q}\,;S;S^{\,\prime}_z,S_z) \times \vec{q} \, \right]}{\ell + 1}\, .} \nonumber 
\end{eqnarray}
In the same way one can introduce according to (\ref{rel6}) the spherical electric and magnetic 
multipole moments $E^{\,\ell m} (S)$ and $M^{\,\ell m} (S)$
of a spin-$S$ particle or composite system using its Breit-frame current distribution by:
\begin{eqnarray} \fl
E^{\,\ell m} (S) \; := \;
E^{\,\ell m} (S;S,S) \; , \quad
M^{\,\ell m} (S) \; := \;
M^{\,\ell m} (S;S,S) 
\end{eqnarray}
with 
\begin{eqnarray}  
\fl \eqalign{E^{\,\ell m} (S;S^{\,\prime}_z,S_z) & := \frac{1}{\ell\, !} \sqrt{\frac{2\,\ell +1 }{4\pi}} \!\int \!d^3q \; \delta^3(-\vec{q}\,) \;
b^{\,\ell m} (-i\,\frac{\partial}{\partial\vec{q}}\;) 
\, j_B^{\,0}(\vec{q}\,;S;S^{\,\prime}_z,S_z) \\
M^{\,\ell m} (S;S^{\,\prime}_z,S_z) & := \frac{1}{\ell\, !} \sqrt{\frac{2\,\ell +1 }{4\pi}} \!\int \!d^3q \; \delta^3(-\vec{q}\,) \;
b^{\,\ell m} (-i\,\frac{\partial}{\partial\vec{q}}\;) 
\, \frac{ \displaystyle\frac{\partial}{\partial\vec{q}} \cdot 
\!\left[ \vec{j}_B (\vec{q}\,;S;S^{\,\prime}_z,S_z) \times \vec{q}\, \right]}{\ell + 1}\, .} \nonumber 
\end{eqnarray}
The corresponding definition of electromagnetic multipole moments in configuration space using the $j_B^{\,\mu} (\vec{r}\,;S)$ and the expressions given in the Section \ref{sec2} is obvious.
\subsection{Consistent definition and derivation of electromagnetic formfactors in QFT}
\subsubsection{Spherical multipole moments, polarization matrices and generalized formfactors} 
\ \\[2mm]
In order to obtain a consistent definition of electromagnetic formfactors in 
momentum space it is useful to introduce the known spherical polarization
matrices $T_{LM} (S;S^{\,\prime}_z,S_z)$ and their Hermitian adjoints 
$T^{\,+}_{LM} (S;S^{\,\prime}_z,S_z)$ in terms of the common (real) 
Clebsch-Gordan coefficients
$<jm|j_1\,m_1,j_2\,m_2>$ (see e.g. \cite[p.\ 44]{var1}): 
\begin{eqnarray} 
\fl \eqalign{T_{LM} (S;S^{\,\prime}_z,S_z) & \; := \; \sqrt{\frac{2L+1}{2S+1}} \; 
<SS^{\,\prime}_z|SS_z,TM> \\
T^{\,+}_{LM} (S;S^{\,\prime}_z,S_z) & \; := \; \sqrt{\frac{2L+1}{2S+1}} \; 
(<SS_z|SS^{\,\prime}_z,TM>)^\ast \; \stackrel{!}{=} \;
(T_{LM} (S;S_z,S^{\,\prime}_z))^\ast \\
 & \; (L=0,1,\ldots,2\,S \quad \mbox{and} \quad M=-L,-L+1,\ldots, L)\, .} 
\end{eqnarray}
They form a complete basis on the set of complex $S\times S$-matrices 
fulfilling the following conditions:
\begin{eqnarray} 
\fl \eqalign{\mbox{``} \;\; \Tr\, [T^{\,+}_{LM} (S)\;T_{L^{\,\prime}M^{\,\prime}} (S)] 
\; \; \mbox{''} & = \;
\sum\limits_{S_zS^{\,\prime}_z} \;
T^{\,+}_{LM} (S;S^{\,\prime}_z,S_z) \; T_{L^{\,\prime}M^{\,\prime}}
(S;S_z,S^{\,\prime}_z\,)
\; = \; \delta_{LL^{\,\prime}} \, \delta_{MM^{\,\prime}} \\
\qquad \;\,\qquad T^{\,+}_{LM} (S;S^{\,\prime}_z,S_z) & = \; (-1)^M \; T_{L-M} (S;S^{\,\prime}_z,S_z) \; .} \nonumber 
\end{eqnarray}
It is now possible to expand the spherical electric and magnetic ``multipole
matrix elements'' in terms of the polarization matrices, i.e.:
\begin{eqnarray} \fl \eqalign{E^{\,\ell m} (S;S^{\,\prime}_z,S_z) & = \; 
\sum\limits_{LM} \, T_{LM} (S;S^{\,\prime}_z,S_z)
\left[ \; \sum\limits_{s^{\,\prime}s} \; T^{\,+}_{LM} (S;s^{\,\prime},s) \;
E^{\,\ell m} (S;s,s^{\,\prime}\,) \; \right] \\
M^{\,\ell m} (S;S^{\,\prime}_z,S_z) & = \; 
\sum\limits_{LM} \, T_{LM} (S;S^{\,\prime}_z,S_z)
\left[ \; \sum\limits_{s^{\,\prime}s} \; T^{\,+}_{LM} (S;s^{\,\prime},s) \;
M^{\,\ell m} (S;s,s^{\,\prime}\,) \; \right] \; .} \nonumber
\end{eqnarray}
Inserting the expressions for $E^{\,\ell m}$ and $M^{\,\ell m}$ given in Section \ref{sbsec1} the expansion reads:
\begin{eqnarray} \fl \eqalign{ 
E^{\,\ell m} (S;S^{\,\prime}_z,S_z) = 
\sum\limits_{LM} \; & T_{LM} (S; S^{\,\prime}_z,S_z) 
\frac{1}{\ell\, !} \; \sqrt{\frac{2\,\ell +1 }{4\pi}} \; \;  \int d^3q \;\; \delta^3(-\vec{q}\,) \\
 & 
b^{\,\ell m} (-i\,\frac{\partial}{\partial\vec{q}}\;) \; \left[ \; \sum\limits_{s^{\,\prime}s} \; T^{\,+}_{LM} (S;s^{\,\prime},s) \, j_B^{\,0}(\vec{q}\,;S;s,s^{\,\prime}\,) 
\; \right] \\
M^{\,\ell m} (S;S^{\,\prime}_z,S_z) = 
\sum\limits_{LM} \; &  T_{LM} (S; S^{\,\prime}_z,S_z)  
 \frac{1}{\ell\, !} \; \sqrt{\frac{2\,\ell +1 }{4\pi}} \; \; \int d^3q  \;\; \delta^3(-\vec{q}\,)
  \\
 & 
b^{\,\ell m} (-i\,\frac{\partial}{\partial\vec{q}}\;) \; \left[ \; \sum\limits_{s^{\,\prime}s} \; T^{\,+}_{LM} (S;s^{\,\prime},s) \, \frac{ \displaystyle\frac{\partial}{\partial\vec{q}} \cdot 
\left[ \vec{j}_B (\vec{q}\,;S;s,s^{\,\prime}\,) \times \vec{q} \right]}{\ell + 1}  
\; \right] . } \label{xy1}  
\end{eqnarray}
The traces are proportional to $b_{\,LM} (\vec{q}\, )$, i.e.:
\begin{eqnarray} 
\eqalign{\sum\limits_{s^{\,\prime}s} \; T^{\,+}_{LM} (S;s^{\,\prime},s) \;
\; j_B^{\,0}(\vec{q}\,;S;s,s^{\,\prime}\,) 
 & \;\; \propto \; b_{\,LM} (\vec{q}\, ) \\
 \sum\limits_{s^{\,\prime}s} \; T^{\,+}_{LM} (S;s^{\,\prime},s) \; 
\; \frac{ \displaystyle\frac{\partial}{\partial\vec{q}} \cdot 
\left[ \vec{j}_B (\vec{q}\,;S;s,s^{\,\prime}\,) \times \vec{q} \; \right]}{\ell + 1}  
 & \;\; \propto \; b_{\,LM} (\vec{q}\, ) \; . } \nonumber
\end{eqnarray}
Now the following properties (see (\ref{usf1})) of the polynomials 
$b^{\,\ell m} (\vec{q}\, )$ will be used:
\begin{eqnarray} 
\eqalign{ b^{\,\ell \,m} (\, - \,i\,\frac{\partial}{\partial\,\vec{q}}\,\, ) \; 
b_{\,\ell \,m^{\,\prime}} (\vec{q}\, ) & = \; (-i\,)^{\,\ell} \; \frac{\ell\,!\,(2\,\ell
\,)!}{2^{\,\ell}} \;\; \delta_{m\,m^{\,\prime}} \\
b^{\,\ell \,m} (\, - \,i\,\frac{\partial}{\partial\,\vec{q}}\,\, ) \; 
b_{\,\ell^{\,\prime} m^{\,\prime}} (\vec{q}\, ) & = \; 
0 \quad \mbox{for} \quad \ell > \ell^{\,\prime} \, .} 
\end{eqnarray}
Observing that all terms with $L > \ell$ in (\ref{xy1}) are proportional to at
least one power of $\vec{q}$ and therefore have to vanish because
of the $\delta$--distribution $\delta^3(-\vec{q}\,)$ the only term surviving in
(\ref{xy1}) is the term with $L=\ell$ and $M=m$, i.e.:
\begin{eqnarray} 
\fl \eqalign{E^{\,\ell m} (S;S^{\,\prime}_z,S_z) \; \stackrel{!}{=} \;
T_{\ell m} (S;S^{\,\prime}_z,S_z) & \;\frac{1}{\ell\, !}  \sqrt{\frac{2\,\ell +1 }{4\pi}} \; \;  \int d^3q \;\; \delta^3(-\vec{q}\,) \;\;
b^{\,\ell m} (-i\,\frac{\partial}{\partial\vec{q}}\;) \\
 & \; \left[ \; \sum\limits_{s^{\,\prime}s} \; T^{\,+}_{\ell m} (S;s^{\,\prime},s) \;
\; j_B^{\,0}(\vec{q}\,;S;s,s^{\,\prime}\,) 
\; \right] \\
 M^{\,\ell m} (S;S^{\,\prime}_z,S_z) \; \stackrel{!}{=} \;
 T_{\ell m} (S;S^{\,\prime}_z,S_z) & \; \frac{1}{\ell\, !} \; \sqrt{\frac{2\,\ell +1 }{4\pi}} \; \; \int d^3q \;\; 
 \delta^3(-\vec{q}\,) \;\;
b^{\,\ell m} (-i\,\frac{\partial}{\partial\vec{q}}\;) \\
 & \; \left[ \; \sum\limits_{s^{\,\prime}s} \; T^{\,+}_{\ell m} (S;s^{\,\prime},s) \; 
\; \frac{ \displaystyle\frac{\partial}{\partial\vec{q}} \cdot 
\left[ \vec{j}_B (\vec{q}\,;S;s,s^{\,\prime}\,) \times \vec{q} \, \right]}{\ell + 1}  
\; \right] \, . } \nonumber 
\end{eqnarray}
It is straight forward to read off generalized electric and magnetic multipole formfactors from these relations, which yield the correct multipole moments.
After separating the dimensionful factor $e/M^\ell$ the following generalized electric and magnetic formfactors   
$F_E^{\,\ell m} (q^2;S;S^{\,\prime}_z,S_z)$ and $F_M^{\,\ell m} (q^2;S;S^{\,\prime}_z,S_z)$ can be defined {\em using 
the Breit--frame charge distribution} ($q^2\stackrel{!}{=}-\,\vec{q}^{\,\,2}$) of a particle or composite system of mass $M$:
\begin{eqnarray} \fl \eqalign{ 
\frac{e}{M^\ell} \, F_E^{\,\ell m} (q^2;S;S^{\,\prime}_z,S_z) \; := \;
T_{\ell m} (S; & S^{\,\prime}_z,S_z) \; b^{\,\ell m} (-i\,\frac{\partial}{\partial\vec{q}}\;) \\
 & \left[ \; \sum\limits_{s^{\,\prime}s} \; T^{\,+}_{\ell m} (S;s^{\,\prime},s) \;
\; j_B^{\,0}(\vec{q}\,;S;s,s^{\,\prime}\,) 
\; \right] \\
\frac{e}{M^\ell} \, F_M^{\,\ell m} (q^2;S;S^{\,\prime}_z,S_z) \; := \;
 T_{\ell m} (S; & S^{\,\prime}_z,S_z) \;
b^{\,\ell m} (-i\,\frac{\partial}{\partial\vec{q}}\;) \\
 & \left[ \; \sum\limits_{s^{\,\prime}s} \; T^{\,+}_{\ell m} (S;s^{\,\prime},s) \; 
\; \frac{ \displaystyle\frac{\partial}{\partial\vec{q}} \cdot 
\left[ \vec{j}_B (\vec{q}\,;S;s,s^{\,\prime}\,) \times \vec{q}\, \right]}{\ell + 1}  
\; \right] \, .} \nonumber 
\end{eqnarray}
Even carrying the angular momentum quantum numbers $\ell$ and $m$, so that one would like to call them {\em spherical}, the generalized formfactors are constructed in such a way, that they yield {\em Cartesian} multipole moments at $q^2=0$.
The connection of the spherical electric and magnetic multipole moments and the
generalized electric and magnetic formfactors is:
\begin{eqnarray} \fl \eqalign{E^{\,\ell m} (S;S^{\,\prime}_z,S_z) & = \; 
\frac{1}{\ell\, !} \; \sqrt{\frac{2\,\ell +1 }{4\pi}} \; \;  \int d^3q \;\; \delta^3(-\vec{q}\,) \;\;
\frac{e}{M^\ell} \; F_E^{\,\ell m} (-\,\vec{q}^{\,\,2};S;S^{\,\prime}_z,S_z) \\
 & \stackrel{!}{=} \; 
\frac{1}{\ell\, !} \; \sqrt{\frac{2\,\ell +1 }{4\pi}} \; \;
\frac{e}{M^\ell} \; F_E^{\,\ell m} (0;S;S^{\,\prime}_z,S_z) \\
M^{\,\ell m} (S;S^{\,\prime}_z,S_z) & = \; 
\frac{1}{\ell\, !} \; \sqrt{\frac{2\,\ell +1 }{4\pi}} \; \;  \int d^3q \;\; \delta^3(-\vec{q}\,) \;\;
\frac{e}{M^\ell} \; F_M^{\,\ell m} (-\,\vec{q}^{\,\,2};S;S^{\,\prime}_z,S_z) \\
 & \stackrel{!}{=} \; 
\frac{1}{\ell\, !} \; \sqrt{\frac{2\,\ell +1 }{4\pi}} \; \;
\frac{e}{M^\ell} \; F_M^{\,\ell m} (0;S;S^{\,\prime}_z,S_z)
\, .} \nonumber 
\end{eqnarray}
From the properties of the Clebsch-Gordan coefficients it is easy to deduce that
the diagonal elements $T_{\ell m} (S;S_z,S_z)$ of the spherical polarization matrices 
are proportional to $\delta_{m0}$ (unveiling slightly the effect of the underlying Wigner-Eckart theorem). Of course this proportionality is also valid
for the generalized electric and magnetic formfactors 
$F_E^{\,\ell m} (q^2;S;S^{\,\prime}_z,S_z)$ and $F_M^{\,\ell m}
(q^2;S;S^{\,\prime}_z,S_z)$, i.e.:
\begin{eqnarray} 
\eqalign{F_E^{\,\ell m} (q^2;S;S_z,S_z) & \stackrel{!}{=} \;
\delta^{\,m0} \; F_E^{\,\ell \,0} (q^2;S;S_z,S_z) \\ 
F_M^{\,\ell m} (q^2;S;S_z,S_z) & \stackrel{!}{=} \;
\delta^{\,m0} \; F_M^{\,\ell \,0} (q^2;S;S_z,S_z) \, .} \nonumber 
\end{eqnarray}
The diagonal case $S^{\,\prime}_z=S_z$ describes the electromagnetic properties of a spin-$S$ particle or composite system in quantum state of spin-polarization $S_z$. It is therefore natural to define the electromagnetic formfactors and multipole moments for a particle or composite system with spin $S$ in the state of maximum spin-polarization, i.e. for the case $S_z = S$ (see e.g.\ Appendix A in \cite{kim1}). Hence the electric and magnetic formfactors 
$F^{\,E\,\ell} (\vec{q}^{\,\,2};S)$ and $F^{\,M\,\ell} (\vec{q}^{\,\,2};S)$ of 
a particle or composite system with spin $S$ can be introduced by:
\begin{eqnarray} 
\fl\eqalign{F_E^{\,\ell m} (q^2;S;S,S) & \stackrel{!}{=} \;
\delta^{\,m0} \; F_E^{\,\ell \,0} (q^2;S;S,S) \; =: \;
\delta^{\,m0} \; F^{\,E\,\ell} (q^2;S)  \\
F_M^{\,\ell m} (q^2;S;S,S) & \stackrel{!}{=} \;
\delta^{\,m0} \; F_M^{\,\ell \,0} (q^2;S;S,S) \; =: \;
\delta^{\,m0} \; F^{\,M\,\ell} (q^2;S) \, .}  
\end{eqnarray}
\subsubsection{Electromagnetic (onshell) formfactors and multipole moments} 
\ \\[2mm]
As a result of the previous considerations the electric 
and magnetic formfactors $F^{\,E\,\ell} (q^2;S)$ and $F^{\,M\,\ell}
(q^2;S)$ of a particle or a composite system of spin $S$ and mass $M$ are defined in momentum space by ($q^2\stackrel{!}{=}-\,\vec{q}^{\,\,2}$):
\begin{eqnarray}  
\fl \eqalign{
\frac{e}{M^\ell} \, F^{\,E\,\ell} (q^2;S) \; := \;
T_{\ell \,0} (S; S,S) \; & b^{\,\ell \,0} (-i\,\frac{\partial}{\partial\vec{q}}\;) \\
 & \left[ \; \sum\limits_{s^{\,\prime}s} \; T^{\,+}_{\ell \,0} (S;s^{\,\prime},s) \;
\; j_B^{\,0}(\vec{q}\,;S;s,s^{\,\prime}\,) 
\; \right] \\
\frac{e}{M^\ell} \, F^{\,M\,\ell} (q^2;S)  \; := \;
 T_{\ell \,0} (S; S,S) \; &
b^{\,\ell \,0} (-i\,\frac{\partial}{\partial\vec{q}}\;) \\
 & \left[ \; \sum\limits_{s^{\,\prime}s} \; T^{\,+}_{\ell \,0} (S;s^{\,\prime},s) \; 
\; \frac{ \displaystyle\frac{\partial}{\partial\vec{q}} \cdot 
\left[ \vec{j}_B (\vec{q}\,;S;s,s^{\,\prime}\,) \times \vec{q} \,\right]}{\ell + 1}  
\; \right] \, .\\
 & } \label{ffak1} 
\end{eqnarray}
For completeness the factors $T_{\ell \, 0} (S;S,S)$ are evaluated to be:
\begin{eqnarray} 
\eqalign{T_{LM} (S;S,S) \;
& = \; \delta_{M\,0} \;\;
\sqrt{\frac{2\,L+1}{2\,S+1}} \;\; \sqrt{\frac{(2\,S+1)\,!\,(2\,S)\,!}{
(2\,S+L+1)\,!\,(2\,S-L)\,!}} \\
 & = \; \delta_{M\,0} \;\; (2\,S)\,! \;\;
\sqrt{\frac{2\,L+1}{
(2\,S+L+1)\,!\,(2\,S-L)\,!}}\;\; .} \nonumber 
\end{eqnarray}
The electric and magnetic formfactors are constructed such, that their values at $q^2 = 0$ are the respective Cartesian electric and magnetic multipole moments, i.e.:
\begin{eqnarray} 
\eqalign{Q_{\,E}^{\,\overbrace{z\ldots z}^{\mbox{\footnotesize $\ell$ times}}} (S)  & \stackrel{!}{=} \;  
\frac{e}{M^\ell} \; F^{\,E\,\ell} (0;S) \\
Q_{\,M}^{\,\overbrace{z\ldots z}^{\mbox{\footnotesize $\ell$ times}}} (S)  & \stackrel{!}{=} \;  
\frac{e}{M^\ell} \; F^{\,M\,\ell} (0;S) \; .}  
\end{eqnarray}
The corresponding expressions for the spherical electric and magnetic multipole
moments are (see e.g.\ (\ref{rel3})):
\begin{eqnarray} 
\eqalign{E^{\,\ell m} (S)  & \stackrel{!}{=} \; \delta^{\,m\,0} \; 
\frac{1}{\ell\, !} \; \sqrt{\frac{2\,\ell +1 }{4\pi}} \; \;
\frac{e}{M^\ell} \; F^{\,E\,\ell} (0;S) \\
M^{\,\ell m} (S)  & \stackrel{!}{=} \; \delta^{\,m\,0} \; 
\frac{1}{\ell\, !} \; \sqrt{\frac{2\,\ell +1 }{4\pi}} \; \;
\frac{e}{M^\ell} \; F^{\,M\,\ell} (0;S) \; .}  
\end{eqnarray}
Returning to the correct description of spin--1 systems in Section~\ref{sec1} one can give now instead of (\ref{idad1}) the correct defining equations for their electromagnetic formfactors, i.e.:
\begin{eqnarray} \fl F_{\,C} (q^2) :=  F^{\,E\,0} (q^2;1) \; , \quad  F_{\,M} (q^2) :=  F^{\,M\,1} (q^2;1) \; , \quad  F_{\,Q} (q^2) :=  F^{\,E\,2} (q^2;1) 
\end{eqnarray}
\section{Summary and outlook} \label{sec4}
Throughout the work presented momentum space expressions for the electromagnetic (onshell) formfactors of a particle or composite system with spin $S$ have been derived (see (\ref{ffak1})) within the Breit-frame which have various appealing properties. The determination of the formfactors, which requires the knowledge of all Breit--frame matrix elements of the current distribution operator, is not based on the specific analytic structure of the underlying current distribution. The formfactors are obtained within the Breit-frame by applying momentum derivatives to the trace of the product of polarization matrices and the matrix constructed by matrix elements of the respective current distribution operator. As a consequence the determination procedure is suitable for not only an analytical, but also a numerical treatment on modern fast computers, which are able to handle numerical derivatives in a reasonable time. The construction of expressions (\ref{ffak1}) ensures that the momentum space expressions for the formfactors are consistent with the classical definition of their counterparts in configuration space, which is not guaranteed in many other theoretical approaches available. The definition of the polynomials $b^{\,\ell m}$ and $c^{\,i_1 \ldots i_\ell}$ yields unambiguous relations between spherical and Cartesian quantities being free of sign ambiguities, which may also be used in polarization physics, i.e.\ spin physics, to get a clear relation between spherical and Cartesian polarization operators.   

Certainly, as the derivations presented yield observable (onshell) quantities related to (transverse) onshell photons --- which is the case for electromagnetic current distributions, formfactors and multipole moments ---, the approach for the moment is not able to give a defining procedure for offshell quantities like offshell formfactors based on longitudinal current distributions, which play an important role in the scattering of virtual photons on particles or composite systems, yet such defining procedures will have to go along a similar line as presented for onshell quantities. For this reason one can raise the question, whether one is able to extract from measurements of differential cross sections of virtual photon scattering (e.g.\ by scattering electrons on particles or composite systems) directly quantities, which one would obtain by the scattering of onshell photons. Similarily one could ask the question, whether on is able to remove the questionmarks on the equal signs of the identifying relations (\ref{idad1}), (\ref{idad2}) and (\ref{idad3}) even after inclusion of all momentum derivatives discussed to the defining equations for the electromagnetic formfactors. The answer of this questions will be the scope of future work. 
Yet, whenever one is able to construct the electromagnetic current distribution of a system within the Breit--frame, then relations (\ref{ffak1}) give a prescription, how to extract the respective momentum space electromagnetic formfactors and related multipole moments consistently. The results presented have been extensively used for the calculations sketched in \cite{kle1} which will be discussed in much more detail in a forthcoming publication.  
\ack
The author is very thankful to the warm hospitality and working atmosphere at the CFIF and the physics department of the Coimbra university, critical remarks, stimulating discussions and support on related topics by --- among others --- M. Dillig, W. Eyrich, W.\ Haeberli, G.\ Rupp, E.\ Steffens and E.\ van Beveren.
This work has been supported by the {\em Funda\c{c}\~{a}o para a Ci\^{e}ncia e a Tecnologia} (FCT) of the {\em Minist\'{e}rio da Ci\^{e}ncia e da Tecnologia} of Portugal under Grant No.\ PRAXIS XXI/BPD/20186/99.
\begin{appendix}
\section{Breit--frame currents for systems of spin 0, 1/2 and 1} \label{appd1}
The momentum space (onshell) matrix elements of electromagnetic current distributions of particles or composite systems with spin 0,
1/2 and 1 within the Breit-frame can be expressed in terms of electromagnetic formfactors in the following way \cite{gla1} ($\vec{q} \stackrel{!}{=} 2\, \vec{k}$, $n^\mu\,:=\,g^\mu_0\,=\,(1,\vec{0}\,)$, $\{\gamma^{\,\mu}, \gamma^{\,\nu}\} = 2\, g^{\,\mu\nu}$, $\sigma^{\,\mu\nu} := i\,[\gamma^{\,\mu}, \gamma^{\,\nu}]/2$, $\bar{u}\, (\vec{p} , s^\prime ) \; u \, (\vec{p} , s) = 2\, M \; \delta_{s^\prime s}$, $u \, (- \, \vec{p} , s) \stackrel{!}{=} \gamma_{\,0} \; u \, (\vec{p} , s) \stackrel{!}{=} \not{\!n}\, u \, (\vec{p} , s)$):\\
\begin{eqnarray} 
\fl \eqalign{j_B^\mu & (\vec{q}\; ; 0 \,; \,0 , 0) \; = \; 
e \; n^\mu \; F_0 (q^2) \\
 & \\
j_B^\mu & (\vec{q}\; ; \frac{1}{2}\,;\,S^{\,\prime}_z\,,\,S_z)
\; = \\
 & = \; 
\frac{e}{(2\,\omega (|\vec{k}\,|))^2} \; \bar{u} (\vec{k},S^{\,\prime}_z)
 \left[ \, 2\,\omega (|\vec{k}\,|) \; n^\mu \; F_0 (q^2) 
 \; + \; i \, \sigma^{\mu\nu}\,q_\nu \; G_1 (q^2)
 \,\right] \, u (-\vec{k}\,,S_z) \\
 & \stackrel{!}{=} \; 
\frac{2\, M\, e}{(2\,\omega (|\vec{k}\,|))^2} \; \bar{u} (\vec{k},S^{\,\prime}_z)
 \left[ \, \gamma^\mu \, F_0 (q^2) 
 \; + \; \frac{i \,\sigma^{\mu\nu}\,q_\nu}{2\,M} \; (G_1 (q^2)-F_0 (q^2))
 \,\right] \, u (-\vec{k}\,,S_z) \\
 & \\
 j_B^\mu & (\vec{q}\; ; 1\,;\,S^{\,\prime}_z\,,\,S_z) 
\; = \\
 & = \; 
 - \; \frac{e}{2\,\omega (|\vec{k}\,|)} \; \left(\varepsilon^{\,\rho \, S_z^{\prime}} (\vec{k}\,)\right)^\ast
 \Bigg[ \, 2\,\omega (|\vec{k}\,|) \; n^\mu  \, 
 \left[ \, g_{\rho\sigma} \, F_0 (q^2) 
  - \, \frac{(q_\rho \,q_\sigma - \frac{1}{3} \,q^2 \,g_{\rho\sigma})}{2\,M^2}
 \; F_2 (q^2)
 \,\right] \\
 & 
 + \, \delta^{\,\mu\nu}_{\rho\sigma}\, q_\nu \; G_1 (q^2)
 \; + \; \frac{i}{M^2} \, [\, q^\mu \,q_\rho \,q_\sigma - 
 \frac{1}{2} \,q^2 \, ( g^{\,\mu}_{\rho} \, q_\sigma + g^{\,\mu}_{\sigma} \,
 q_\rho)]
 \; G_2 (q^2)
 \Bigg] \, \varepsilon^{\,\sigma \, S_z} (-\vec{k}\,) \; .} \nonumber
\end{eqnarray}
\end{appendix}


\section*{References}
\begin{thebibliography}{99}
\bibitem{gla1} Glaser V and Jak\v{s}i\'{c} B 1957 {\it Il} \NC {\bf V} 1197
\bibitem{ohl1} Ohlsen G G 1972 \RPP {\bf 35} 717
\bibitem{sta1} Stapp H P 1956 \PR {\bf 103} 425
\bibitem{bas1} 1960 {\it Proc.\ Int.\ Symp.\ on Polarization Phenomena of Nucleons  (Basel)} \\ 1960 {\it Helv.\ Phys.\ Acta Supp.} {\bf VI} 436
\bibitem{mad1} 1970 {\it Proc.\ 3rd Int.\ Symp.\ on Polarization Phenomena in Nuclear Reactions  (Madison, Wisc.)} (Madison: The University of Wisconsin Press, 1971) p~xxv
\bibitem{hon1} Honzawa N and Ishida S 1992 \PR C {\bf 45} 47
\bibitem{gou1} Gourdin M 1963 {\it Il} \NC {\bf XXVIII} 533
\bibitem{bro2} Brodsky S J and Hiller J R 1992 \PR D {\bf 46} 2141
\bibitem{zui1} Zuilhof M J and Tjon J A 1980 \PR C {\bf 22} 2369
\bibitem{kim1} Kim K J and Tsai Y-S 1973 \PR D {\bf 7} 3710
\bibitem{are1} Arenh\"ovel H, Ritz F and Wilbois T 2000 \PR C {\bf 61} 034002
\bibitem{are2} Arenh\"ovel H and Singh S K 2000 {\it Preprint} nucl-ph/0012066
\bibitem{akh1} Akhiezer A I, Sitenko A G and Tartakovskii V K 1994 {\it Nuclear electrodynamics} (Springer-Verlag Berlin Heidelberg, 1994) 
\bibitem{hon2} Honzawa N and Ishida S 1985 {\it Prog.\ Theor.\ Phys.} {\bf 74} 939
\bibitem{gar1} Gar\c{c}on M \etal 1994 \PR C {\bf 49} 2516
\bibitem{kle1} Kleefeld F 2000 {\it Proc.\ XVII'th European Conf.\ on ``Few-Body Problems in Physics'' (\'{E}vora, Portugal)}  
               to be published in 2001 \NP A \\
               (Kleefeld F 2000 {\it Preprint} nucl-ph/0010002)
\bibitem{bro1} Bronzan J B 1971 {\it American J.\ Phys.} {\bf 39} 1357
\bibitem{var1} Varshalovich D A, Moskalev A N and Khersonskii V K 1988 {\it Quantum theory of angular momentum} (World Scientific Publishing Co.\ Pte.\ Ltd., 1988) 
\bibitem{gro5} Gross F and Riska D O 1987 \PR C {\bf 36} 1928
\end{thebibliography}
\end{document}